\documentclass[12pt]{iopart}

\usepackage{graphicx}
\usepackage{siunitx}

\begin{document}

\title[]{Oscillations of a suspended slinky}

\author{J\"org Pretz}

	\address{Institut f\"ur Kernphysik, Forschungszentrum J\"ulich, 52425 J\"ulich, Germany \\
	III. Physikalisches Institut B, RWTH Aachen University, 52056 Aachen, Germany \\
	JARA-FAME, Forschungszentrum J\"ulich und RWTH Aachen University }

\ead{pretz@physik.rwth-aachen.de}
\vspace{10pt}
\begin{indented}
\item[]May 2020
\end{indented}

\begin{abstract}
This paper discusses the oscillations of a spring (slinky) under its own weight. 
A discrete model, describing the slinky 
by $N$ springs and $N$ masses, is introduced and 
compared to a continuous treatment.
One interesting result is that the upper part of the slinky
performs a triangular oscillation whereas the bottom part
performs an almost harmonic oscillation if the slinky starts
with "natural" initial 
conditions, where the spring is just pulled further down from its rest 
position under gravity and then released.

It is also shown that the period of the oscillation 
is simply given by $T=\sqrt{32 L/g}$, where $L$ is the length
of the slinky under its own weight and $g$ the acceleration of gravity independent of the other properties of the spring.
\end{abstract}

%

\submitto{\EJP}
%
%
%

\section{Introduction}
A slinky, invented in the 1940 by Richard James, is in the
context of this paper a spring that
is oscillates under its own weight without any additional mass
attached to it with a quality factor high enough to observe the
oscillations.
There are many articles on a falling slinky~\cite{Cross,Unruh,Vanderbei} and
the suspended slinky~\cite{Bowen,Young}. 
This article studies interesting aspects of a suspended slinky.
Equations of motion are derived and solved for the discrete case where the slinky is described by $N$ masses and springs (section~\ref{sec:discrete}).
Section~\ref{sec:cont} treats the continuous case.
Section~\ref{sec:exp} compares the results obtained analytically to experimental results.

\section{Discrete case}\label{sec:discrete}
\begin{figure}
	\begin{center}
	\includegraphics[width=0.5\textwidth]{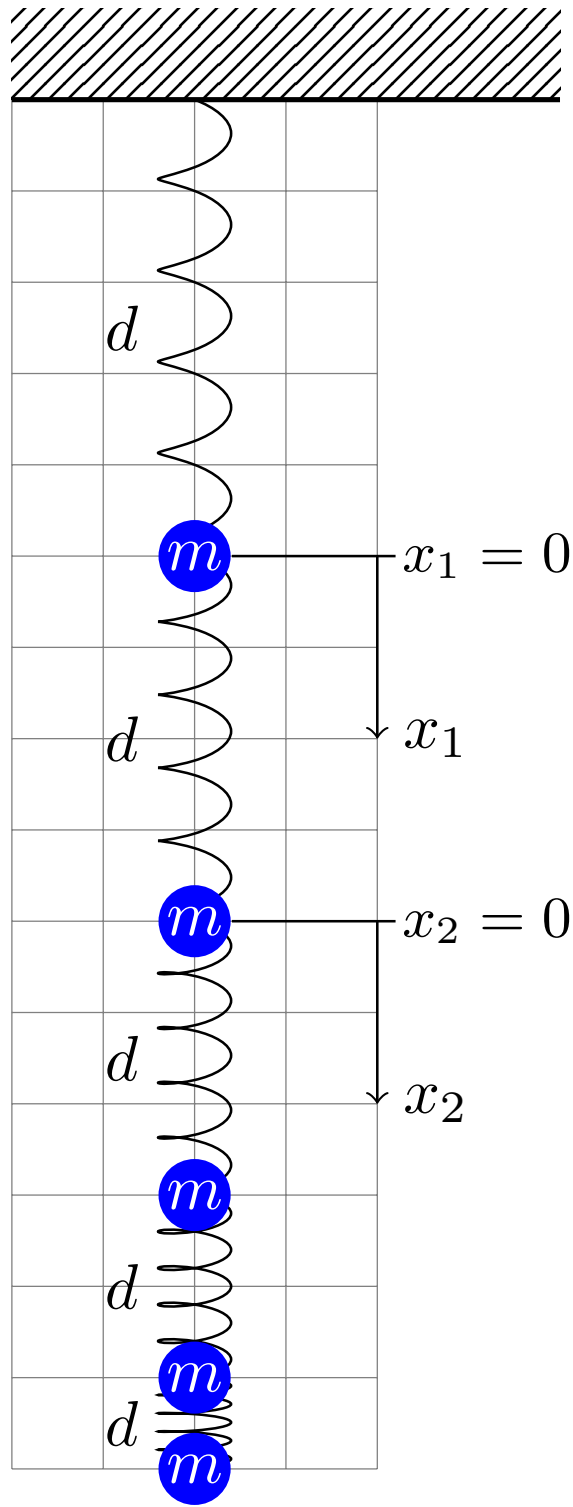}
	\end{center}
\caption{Slinky modeled as $N$ masses and $N$ springs. The picture shows the rest position under its own weight.\label{fig:slinky_pendel}}
\end{figure}

The slinky is modeled by $N$ identical massless springs with spring constant $d$ and mass $m$
as shown in Figure~\ref{fig:slinky_pendel}.
A given mass experiences forces from the two neighboring springs
leading to the following equation of motion
\begin{equation}
  m \ddot{x}_j = -d (x_j - x_{j-1}) - d (x_j - x_{j+1}) \, ,
\end{equation}
where the $x_j$ denotes the excursion from the rest position of the mass. An exception is the first and last spring.
In this case the equations of motion read
\begin{eqnarray}
  m \ddot{x}_1 &=& -d x_1 - d (x_1-x_2) \, ,\\ 	
  m \ddot{x}_N &=& -d (x_N -x_{N-1}) \, .
\end{eqnarray}
This results in the  following system of coupled differential equations
\begin{equation}\label{eq:eom}
\ddot{\bi{x}} + \omega_0^2 B \bi x  = 0   \, ,
\end{equation}
with
	\begin{equation}
B=\left(
\begin{array}{rrrrrr}
2 & -1 & 0 & 0 &\dots & 0   \\
-1 & 2 & -1 & 0 & \dots & 0\\
0 & -1  & 2 & -1 & \dots & 0\\
\vdots &  & &\ddots & & \vdots \\
0 & \dots  & 0  & -1 &  2 &  -1 \\
0 & \dots  &  0 &  0 &   -1 & 1 \\ 
\end{array}
\right) \, 
\end{equation}
and 
\begin{equation}
\omega_0^2 = d/m \,  .
\end{equation}

In contrast to the corresponding matrix of a falling slinky ~\cite{Vanderbei}, there is a $2$ in the upper left corner instead of a 
$1$.

A solution with initial condition $\dot {\bi x}(0) = 0$ is:
\begin{equation}\label{eq:xt}
\bi x(t) = \bi a \, \cos( \omega t)  \, .
\end{equation}

Inserting equation~\ref{eq:xt} into equation~\ref{eq:eom}
leads to the eigenvalue problem
\begin{equation}
-\omega^2 \bi a + \omega_0^2 B \bi a = 0 \, .
\end{equation}


The general solution of equation~\ref{eq:eom} is thus
\begin{equation}\label{eq:xt1}
\bi x(t) = \sum_{j=0}^N c_j \bi a_j \, \cos(j\omega_{j} t)  \, ,
\end{equation}  
where $\omega_j = \sqrt{|\omega_j^2|}$ are the the square roots of the eigenvalues
and $\bi{a}_j$ the corresponding eigenvectors.
For a given initial condition $\bi x(0) = \bi x_0$,
the vector $\bi c$ is determined by inverting
\begin{equation}\label{eq:x0}
  \bi{x}_0 = A \bi c \, ,
  \end{equation}
where $A$ is the matrix with the eigenvectors as columns.   

We assume the springs to be massless and of zero length when not 
exposed to a force.
Then the rest position of the first (top) mass is given by
\begin{equation}
(\bi{x}_{\mathrm{rest}})_{1} = N \frac{mg}{d} \, .
\end{equation}
The first spring is stretched by all $N$ masses.
The second spring is accordingly stretched by $N-1$ masses.
This leads to 
\begin{eqnarray}
\bi{x}_{\mathrm{rest}} &=&  \,  \frac{mg}{d} \, (N, N+(N-1), N+(N-1)+(N-2), \dots) \, . \label{eq:xrest}
\end{eqnarray}
The position of the $N$-th mass is thus
\begin{equation}\label{eq:xrestL}
 (\bi{x}_{\mathrm{rest}})_{N} = \frac{mg}{d} \, \frac{N(N+1)}{2} \, = L \, .
\end{equation}

To study oscillations
we just pull the slinky further down from its rest position.
The initial condition (deviation from $\bi x_{\mathrm{rest}}$) is given by
\begin{equation}
  \bi x_0 = X_0 \, \frac{2}{N(N+1)} \, \frac{d}{mg} \,\bi x_{\mathrm{rest}} \, .
\end{equation}
For the first and last mass from one finds:
\[
  (\bi{x}_{0})_{1} = \frac{2}{N+1} X_0 \, , \quad (\bi{x}_{0})_{N} = X_0 \, .
\]

\begin{figure}
  \begin{center}
	\includegraphics[width=\textwidth]{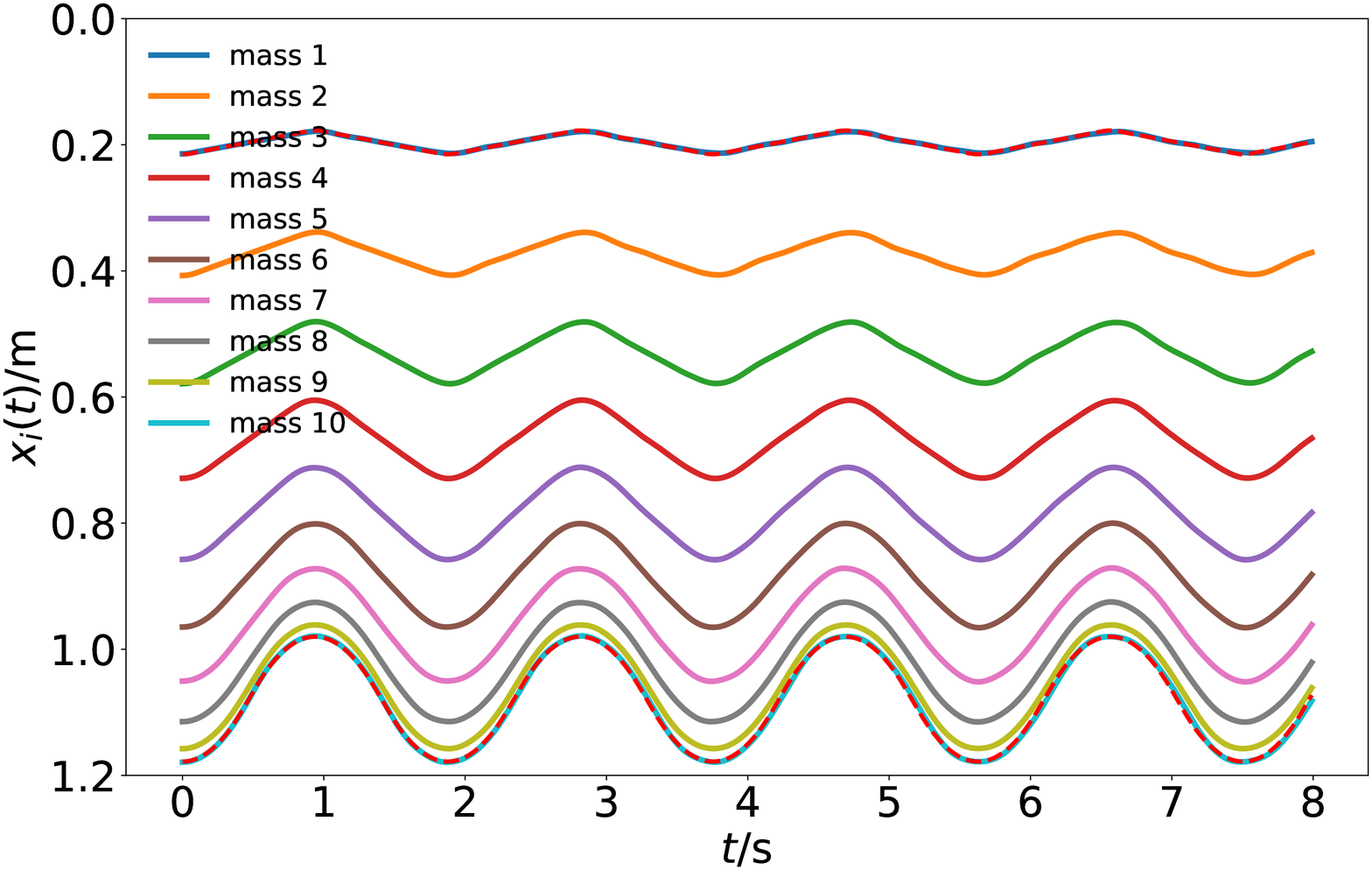}
\end{center}
	\caption{Oscillations as a function of time $t$ for the different masses.\label{fig:slinky_osc}}
\end{figure}

Now that the initial conditions are fixed one can
calculate the oscillations
using equation~\ref{eq:xt1}.
Figure~\ref{fig:slinky_osc} shows the results for the parameters given in table~\ref{tab:para1}.
\begin{table}
\caption{Parameters for oscillations shown in  Figure~\ref{fig:slinky_osc}.
	 \label{tab:para1}}
\begin{indented}
  \lineup
\item[]\begin{tabular}{@{}rll}
  \br
		parameter & value & meaning\\
		\mr
		$N$   &   10   & number of masses\\
		$X_0/\si{m}$ & \00.1  & deviation from rest position for bottom mass\\
			$D /\si{kg \,s^{-2}} $   & \00.15  &spring constant of slinky\\
		$d=N D/\si{kg \, s^{-2}}$    & \01.5  & corresponding spring constant of single spring  \\	
		       &         &  between two masses in Fig.~\ref{fig:slinky_pendel} \\
	$M/ \si{g}$   &  30 & mass of slinky \\
	\br
\end{tabular}
\end{indented}
\end{table}
The amplitudes are the larger
the lower the position along the slinky.
At the top, the oscillations have a triangular shape whereas
at the bottom the oscillation is more sinusoidal.
In the next section we try to understand this behavior 
by going to a continuous distribution of the mass over the slinky.

\section{Continuous case}\label{sec:cont}
If $N\rightarrow \infty$ and $m\rightarrow 0$ such that $Nm = M$ 
remains constant, one reaches a continuous mass distribution.
This system can best be described by a dimensionless variable 
$n$ which is defined by the turn number of the spring divided by
the total number of turns, i.e. $0\le n \le 1$~\cite{Young}.
The position $x$ as a function of $n$ along the slinky under gravity can be derived from the following consideration. 
Under gravity the stretching $x(n+dn) - x(n)$ of the slinky is proportional
to the remaining mass below the position $n$. Since 
this mass is proportional to $(1-n)$, one finds:
\begin{equation}
  x(n+ \mathrm{d}n) - x(n) \propto (1-n) \mathrm{d}n \, .
\end{equation}
This leads to
\begin{equation}\label{eq:x}
  x(n) = (2n - n^2) L \, ,
\end{equation}
normalized such that the total length is $x(1)=L$.
Inverting equation~\ref{eq:x} leads to
\begin{equation}
n(x) = 1- \sqrt{1- \left(\frac{x}{L}\right)} \, .
\end{equation}
\ref{app:weq} shows in detail that, 
using $n$ instead of the vertical position $x$ as a coordinate, the system 
can be described by the following wave
equation~\cite{Young,Gluck}
\begin{equation}\label{eq:waveeq}
\frac{\partial^2 s(n,t)}{\partial t^2} - \frac{g}{2L} \frac{\partial^2 s(n,t)}{\partial n^2} = 0 \, ,
\end{equation}
where a wave is propagating with constant velocity 
$v = \sqrt{\frac{g}{2L}}$. This is not the case if the coordinate
$x$ is used instead of $n$.
$s(n,t)$ is the deviation of the slinky from its rest position
as a function of the relative turn number $n$ and time $t$.

If we pull the slinky at the bottom, according to equation~\ref{eq:x} the initial conditions are  given by
\begin{equation}\label{eq:s0}
   s(n,0) = (2 n- n^2)  X_0   \, ,\\
\end{equation}
where $X_0$ denotes the excursion of the bottom of the slinky
from its rest position.
Further initial and boundary conditions are
\begin{eqnarray}
  \dot s(n,0) &=& 0 \label{eq:dsdt} \, , \\
  s(0,t) &=& 0 \quad \mbox{and}\\
  \frac{\mathrm{d} s(1,t)}{\mathrm{d} n} &=& 0 \, .\label{eq:dsdn}
\end{eqnarray}
Equation~\ref{eq:dsdn} assures an anti-node at the open end $n=1$.

The solution is given by
\begin{equation}\label{eq:snt}
 s(n,t) = \sum_{j=0}^{\infty} A_j \sin( k_j n) \cos(\omega_j t) \, .
\end{equation}
To fulfill condition equation~\ref{eq:dsdn}, we have
\begin{equation}
k_j = (2j+1) \, \frac{\pi}{2} \, , \quad j=0,1,2,\dots .
\end{equation}
Finally equation~\ref{eq:waveeq} leads to:
\begin{equation}
\omega_j = \sqrt{\frac{g}{2L}} \, k_j \, .
\end{equation}
The fundamental frequency is thus given
by 
\[
\omega_0 = \sqrt{\frac{g}{2L}} \, \frac{\pi}{2} \, .
\]
For the period length of the slinky one finds
\begin{equation}\label{eq:T}
T = \sqrt{\frac{32L}{g}} \, ,
\end{equation}
independent of the mass or other properties of the slinky. This result has been derived in~\cite{Young}
in a different context as a round trip time of a pulse.

The length $L$ under its own weight is easily derived
from equation~\ref{eq:xrestL}
\begin{equation}
L =  \lim_{N\rightarrow \infty, m\rightarrow 0}  \frac{mg}{d} \, \frac{N(N+1)}{2}   =\frac{Mg}{2D} \, .
\end{equation}
Note that $d$ is the spring constant of a single spring (see Fig.~\ref{fig:slinky_pendel}). The spring constant of the total spring is $D = d/N$.
The dependence of the period duration $T$
on $M$ and $D$ is hidden in the length $L(M,D)$.
$T$ can also be expressed as
\[
  T = 4 \sqrt{\frac{M}{D} }\, .
\]

\begin{figure}
  	\begin{center}
	\includegraphics[width=\textwidth]{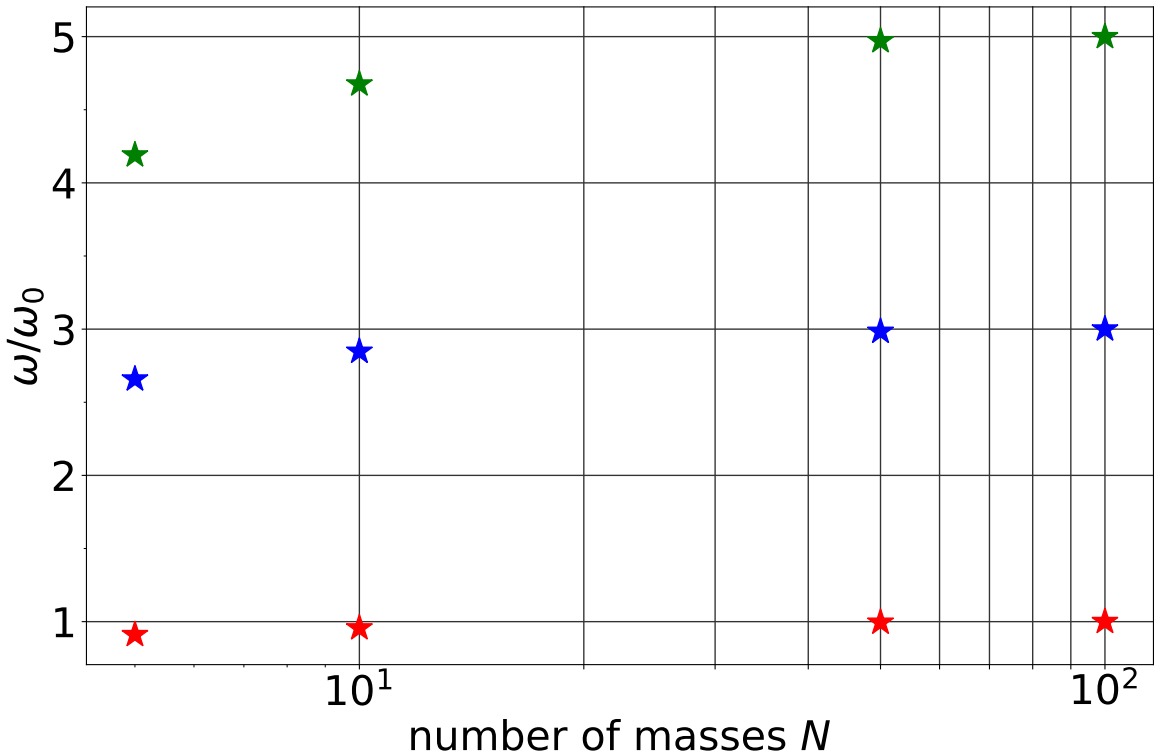}
	\end{center}
	\caption{First three frequencies for $N=5,10,50,100$ divided by the fundamental frequency $\omega_0$ for the continuous case.\label{fig:freq_N}}
\end{figure}

For the parameters given in table~\ref{tab:para1}, one finds
$\omega_0 = 3.49\, \si{s^{-1}}$ which is very close to the solution found in the discrete case for $N=10$,  $\omega_{0, N=10} =3.34 \, \si{s^{-1}}$. Figure~\ref{fig:freq_N} shows a comparison 
of the frequencies for various values of $N$ for the first
three contributing frequencies. 

The coefficients $A_i$ in equation~\ref{eq:snt} are found by a Fourier analysis. As shown in detail in~\ref{app:fourier}, in order
to satisfy the initial condition equation~\ref{eq:s0}, one finds
\[
A_i \propto \frac{32}{\pi^3 (2j + 1)^3} X_0 \, , \quad  j=0,1,2,3, \dots \, .
\]
The amplitude of the different frequencies is given by 
the factors $A_j \sin(k_j n)$ in equation~\ref{eq:snt}.
At small $n$, i.e. at the top of the spring, one finds
\[
 A_j \sin(k_j n) = X_0 \frac{32}{\pi^3 (2j + 1)^3} k_i n = X_0  \frac{16}{\pi^2 (2j+1)^2} n \, .
\]
The amplitudes have the ratios 1:9:25:\dots which corresponds
to a triangular shape as observed in Figure~\ref{fig:slinky_osc}.
At the bottom end of the slinky, i.e. $n=1$ one has $\sin(k_j n)=1$ 
resulting in
\begin{eqnarray}
  A_j \sin(k_j n) &=&   X_0 \frac{32}{\pi^3} \approx \quad 1.032 X_0\,  \, \mbox{for} \quad j=0 \, ,\\
   &=&   X_0 \frac{32}{27 \pi^3 } \approx \quad 0.038 X_0 
   \,  \, \mbox{for}  \quad  j=1 \, , \\
   &=&   X_0 \frac{32}{125 \pi^3 } \approx \quad 0.0083 X_0 \, \mbox{for} \quad  j=2 \, , 
\end{eqnarray} 
which almost corresponds to a pure sine wave.
But even in the limit $N \rightarrow \infty$ the bottom mass
oscillation is not a purely harmonic.

In Fig.~\ref{fig:slinky_osc} the dotted lines for the top
and bottom mass show the solution (at $n=0.1$ and $n=1$ respectively) from the wave equation
which agrees perfectly with the discrete solution with $N=10$.

\section{Comparison to experiment}\label{sec:exp}

\begin{figure}
	\includegraphics[width=\textwidth,viewport=200 0 3900 2100,clip]{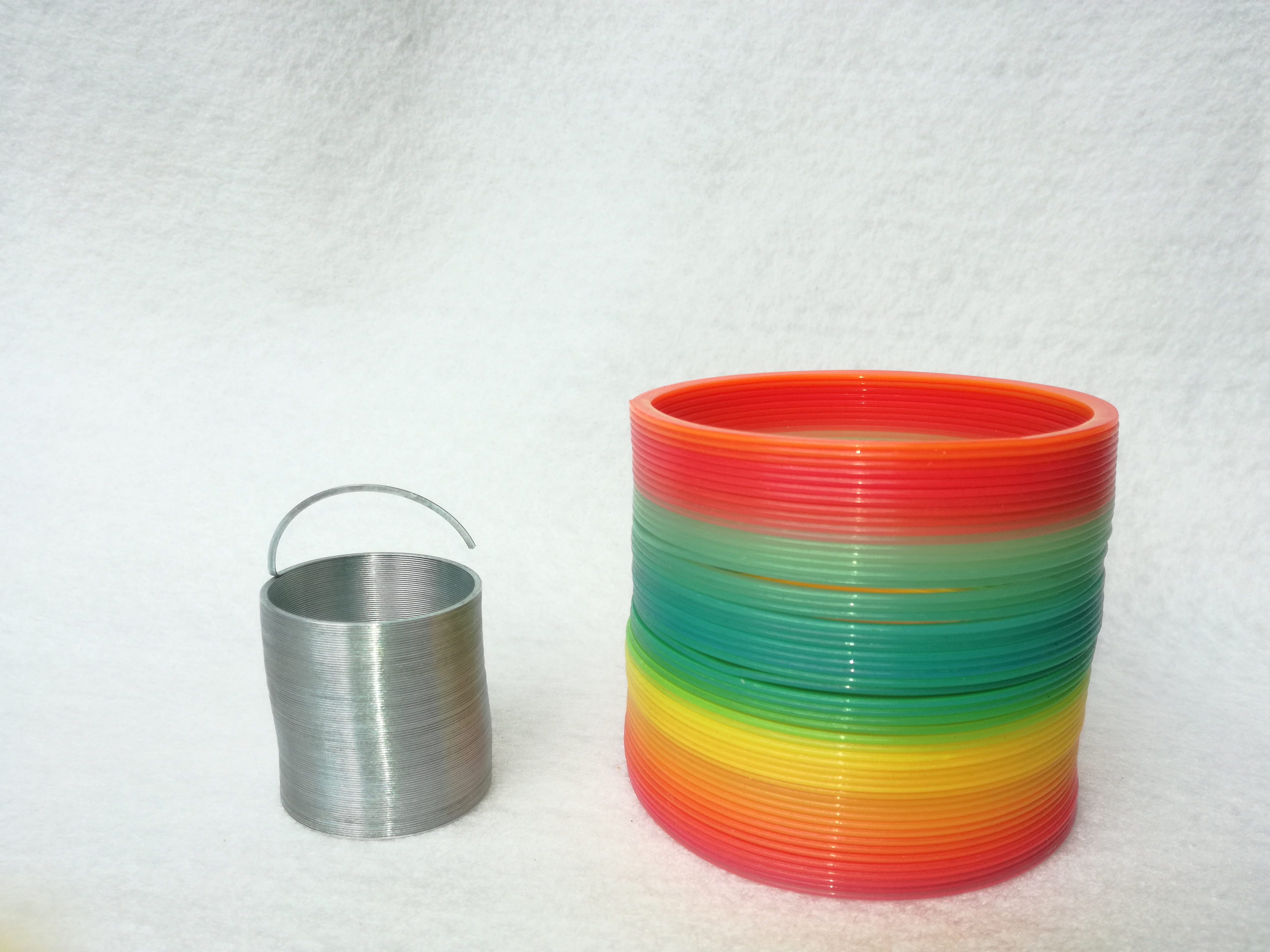}
\caption{Photograph of the two slinkies used in the experiemnts.\label{fig:photo_slinky}}
\end{figure}

Two slinkies (see Figure~\ref{fig:photo_slinky} and Table~\ref{tab:slinkies}) were used
to confirm equation~\ref{eq:T} and to verify the motion
of the top and bottom mass.

\begin{table}
\caption{Parameters of slinkies used in experiments\label{tab:slinkies}.}
\begin{indented}
\item[]\begin{tabular}{@{}lll}
  \br
	&	plastic & metal \\
	\mr
  nb. of turns & 52  &  93 \\
  mass $M/\si{g}$ &65 &  35        \\
  length $L/\si{m}$ under gravity  & 2.40 & 0.32\\
   unstretched length $L_0/\si{m}$   & 0.07 & 0.04 \\
   \br
\end{tabular}
\end{indented}
\end{table}


\begin{table}
\caption{Results of the measurements with the two slinkies. $L_0$ is the length in the unstretched state. 
	The numbers in parentheses indicate the uncertainties. \label{tab:measurements}}
  \begin{indented}
\item[]\begin{tabular}{@{}lllll}
  \br
length $L/\si{m}$ & length $L_0/ \si{m}$ & $T_{meas}/\si{s}$&$T_{theo}/\si{s}$ & \\  
\mr
2.400(5)  &  0.070(2) &  2.50(2) & 2.80(13) & plastic\\     
1.000(5)  &  0.040(2) &  1.65(2) & 1.81(13) &plastic\\
0.320(5)  & 0.040(2) &  1.01(2) & 1.02(13) & metal\\
0.090(5)  &  0.020(2) &  0.59(2) & 0.54(13) & metal\\
\br
\end{tabular}
\end{indented}
\end{table}

\begin{figure}
	\includegraphics[width=\textwidth]{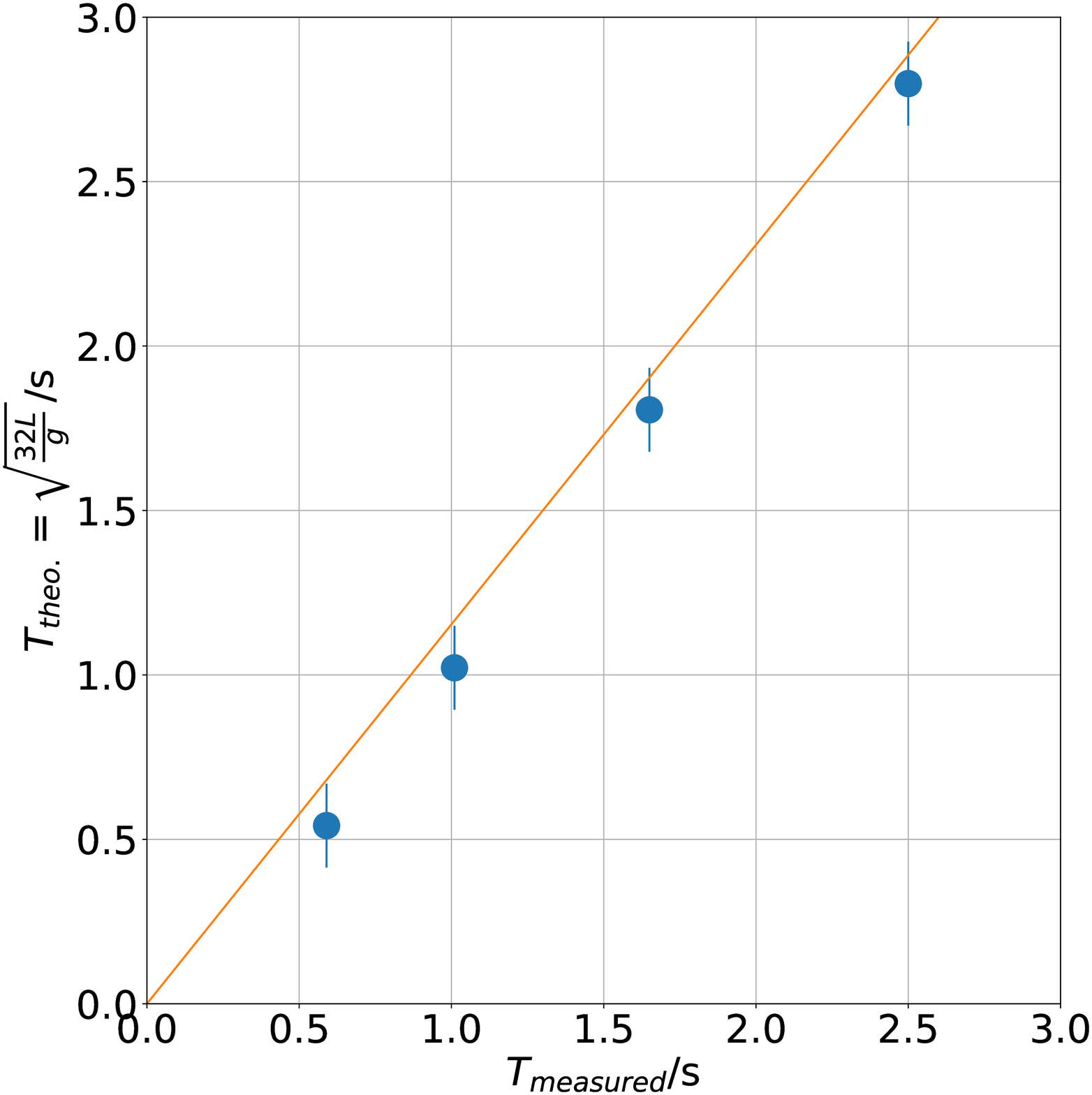}
	\caption{The calculated period length (equation~\ref{eq:T}) vs. the measured period length.\label{fig:period}}
\end{figure}

Figure~\ref{fig:period} shows a comparison of the measured and calculated period duration. A reasonable agreement is found.

\begin{figure}
	\includegraphics[width=\textwidth]{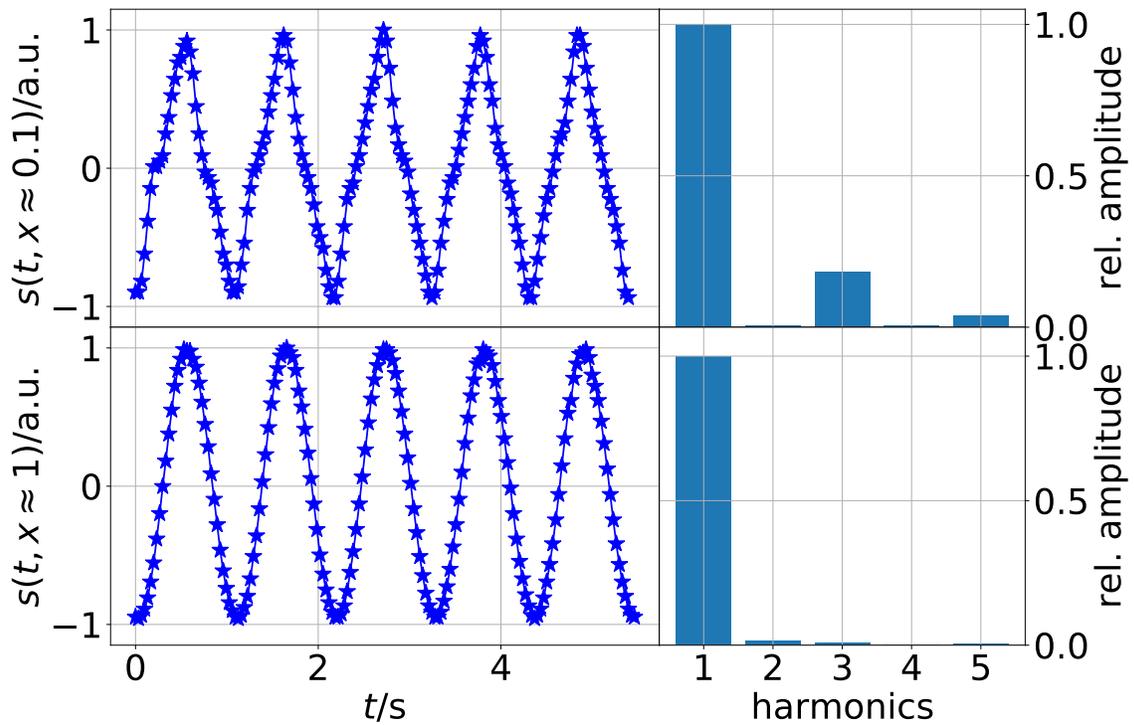}
	\caption{Left: Oscillation of top and bottom part of the slinky. Right: The corresponding frequency spectrum.\label{fig:osc_top_bot}}
\end{figure}

Figure~\ref{fig:osc_top_bot} shows the measurement of an oscillating
slinky. The top plots are the oscillations and frequencies of a point at the top
of the slinky, the bottom plots are the corresponding plots 
for the bottom end of the slinky. 
Qualitatively one observes the behavior in Figure~\ref{fig:slinky_osc}. The oscillation of the top point
is more triangular shaped.  
It is worthwhile to mention that in these experiments it is important
to pull the slinky from the bottom to have the correct initial conditions. If the slinky is for instance pulled at $n\approx 0.4$, the
oscillations look very different. In principle this could be studied
by choosing the corresponding initial conditions in equations~\ref{eq:x0} or \ref{eq:s0}.

\begin{figure}
	\includegraphics[width=\textwidth]{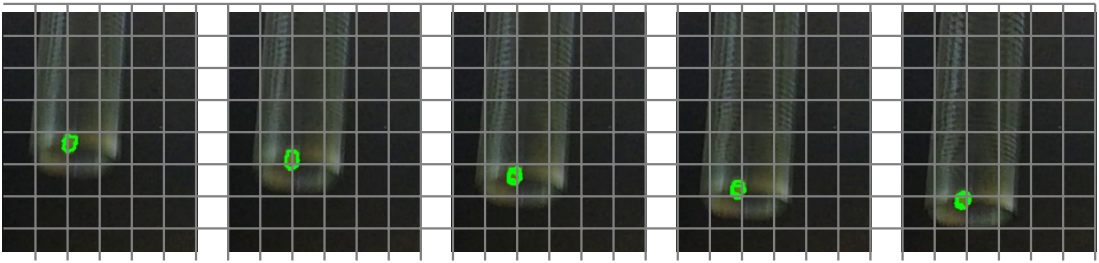}
	\caption{Analysis of five video frames using the opencv library
	to track a point at the bottom of the  slinky~\cite{opencv_library}. The result 
	of the tracking is indicated by the green areas.
	 The images are taken 0.033\,\si{s} apart.\label{fig:tracking}}
\end{figure}

The points of Figure~\ref{fig:osc_top_bot}  were obtained using the open source opencv library~\cite{opencv_library} which allows easily to track a point in a video according to its color.
Figure~\ref{fig:tracking} shows the analysis of five video frames,
where a point at the bottom of the slinky is tracked.

\section{Summary and conclusions}
Starting from discrete treatment with $N$ masses attached to $N$
springs the vertical motion of a suspended slinky was studied.
A first observation of the discrete treatment is that the top part of the slinky performs triangular motions whereas the bottom
part is more sinusoidal. Using a continuous model this could also
be confirmed analytically. Measurements show a good agreement
with the analytic results derived.

\ack
I would like to thank J.~Barth for comments and suggestions on the paper.

\newpage
\appendix

\section{Derivation of the wave equation}\label{app:weq}
The wave equation for a string subject to a force $F$ along
the longitudinal direction reads
\begin{equation}
\frac{\partial^2 s(x,t)}{\partial t^2} - \frac{F}{\rho A} \frac{\partial^2 s(x,t)}{\partial x^2} = 0 \, ,
\end{equation}
where $\rho$ is the density and $A$ the
cross-sectional area of the string.

The force at a given $n$ is given by
\begin{equation}\label{eq:force}
F(n) = (1-n) Mg
\end{equation}
where $M$ is the total mass of the slinky.

From equations~\ref{eq:x} and \ref{eq:force} we can express 
the force and the density $\rho$ as a function of the position $x$.
\begin{eqnarray}
F(x) &=& (1-n(x))Mg =
\sqrt{1-\frac{x}{L}} \, Mg \, ,\\
\rho(x) &=& \frac{\mathrm{d}n}{\mathrm{d}x} \, \frac{M}{A}  = \frac{1}{L} \, \frac{1}{2 \sqrt{1-\left(\frac{x}{L}\right)}} \frac{M}{A} \, .
\end{eqnarray}
This leads to 
\begin{equation}
\frac{\partial^2 s(x,t)}{\partial t^2} - 2 gL\left(1-\frac{x}{L}\right) \frac{\partial^2 s(x,t)}{\partial x^2} = 0 \, .
\end{equation}

As in references~\cite{Young,Gluck} it is easier to write the wave equation
in terms of the turn number $n$
\begin{equation}
\tilde s(n,t) = s(x(n),t) \frac{d x}{d n} = 
s(x(n),t) \, 2L \, \sqrt{1-\left(\frac{x}{L}\right)} \, .
\end{equation}
Using
\[
\frac{\partial}{\partial x} = \frac{\partial }{\partial n}
\frac{d n}{d x } = \frac{\partial }{\partial n} \, 
\left(2 L \, \sqrt{1-\left(\frac{x}{L}\right)}\right)^{-1}
\]
and dropping the tilde on $s$, the wave equation becomes
\begin{equation}\label{eq:waveeq1}
\frac{\partial^2 s(n,t)}{\partial t^2} - \frac{g}{2L} \frac{\partial^2 s(n,t)}{\partial n^2} = 0 \, .
\end{equation}
In terms of the relative turn number $n$ the velocity 
$\mathrm{d}n/\mathrm{d}t = \sqrt{\frac{g}{2L}}$ is constant.

\section{Determination of coefficients $A_i$}\label{app:fourier}

Equations~\ref{eq:s0} and~\ref{eq:snt} for $t=0$ read
\[
   X_0 (2 n - n^2) = \sum_{j=0}^{\infty}   A_j \sin( k_j n)  
\] 
For $j=0$ the sine term performs only one quarter of an oscillation between
$n=0$ and $n=1$. Therefore the following integral
is only evaluated over a quarter period length up to
$\lambda/4$. To get the correct normalization the result
has to be multiplied by 4. This results in the following expression
for the Fourier coefficients $A_i$:
\begin{equation}
  A_i = 4 \int_0^{\lambda/4} \sin(kn) \, (2n-n^2) X_0 \mathrm{d}n \, \frac{k_0}{\pi} = \frac{32}{\pi^3 (2j + 1)^3} X_0 \,
\end{equation}
with $\lambda={2\pi}/{k_0}$.

\clearpage

\bibliography{literature_slinky.bib}

\bibliographystyle{ieeetr}

\end{document}